\documentclass[a4paper,fleqn]{cas-sc}

\usepackage[authoryear]{natbib}
\usepackage{float}
\usepackage{hyperref}
\usepackage{lineno}
\def\tsc#1{\csdef{#1}{\textsc{\lowercase{#1}}\xspace}}
\tsc{WGM}
\tsc{QE}
\tsc{EP}
\tsc{PMS}
\tsc{BEC}
\tsc{DE}

\begin{document}
\let\WriteBookmarks\relax
\def\floatpagepagefraction{1}
\def\textpagefraction{.001}

\shorttitle{Evaluating Airline Service Quality Through the Comprehensive Text-mining and TOPSIS-VIKOR-AISM Analysis}

\shortauthors{H. Xie et~al.}

\title [mode = title]{Evaluating Airline Service Quality Through the Comprehensive Text-mining and TOPSIS-VIKOR-AISM Analysis}                      

\author[1]{Haotian Xie}[type=editor, orcid=0000-0001-6352-1979]

\cormark[1]

\affiliation[1]{organization={Faculty of Arts and Sciences, Beijing Normal University},
    city={Zhuhai},
    postcode={519087}, 
    country={China}}

\ead{haotianxie@mail.bnu.edu.cn}

\credit{Conceptualization, Methodology, Investigation, Software, Data curation, Formal analysis, Writing – original draft, Writing – review \& editing, Validation, Supervision}

\cortext[cor1]{Corresponding author. These authors contributed equally: Haotian Xie, Yi Li.}

\author[1]{Yi Li}[type=editor]
\credit{Conceptualization, Methodology, Formal analysis, Writing – original draft, Writing – review \& editing, Validation}

\author[2]{Yang Pu}[type=editor]
\affiliation[3]{organization={Bay Area International Business School, Beijing Normal University},
    city={Zhuhai},
    postcode={519087}, 
    country={China}}
\credit{Software, Data curation, Formal analysis, Writing – review \& editing, Validation}

\author[1]{Chen Zhang}[type=editor]

\credit{Formal analysis, Writing – review \& editing, Validation}

\author[2]{Junlin Huang}[type=editor]

\credit{Formal analysis, Validation}

\begin{abstract}
Service quality rankings are pivotal for maintaining sustainability in the fiercely competitive airline industry. However, prior research in this domain has often fallen short in aspects of sample size, efficiency, and dependability. This study introduces refined insights into this area and establishes a comprehensive, yet highly elucidative, ranking framework. Initially, we employ Latent Semantic Analysis (LSA) to distill principal themes and sentiments from online reviews of 80 airlines. Subsequently, we utilize the SentiWordNet lexicon and the TextBlob package for conducting sentiment analysis based on these reviews. Following this, we construct a hierarchical structure using the computation of compromise solutions, employing an integrated Technique for Order Preference by Similarity to Ideal Solution, vis-à-vis Kriterijumska Optimizacija I Kompromisno Resenje-Adversarial Interpretive Structural Model (TOPSIS-VIKOR-AISM) methodology. Beyond aiding consumer decision-making and fostering airline growth, this study contributes novel viewpoints on evaluating the efficacy of airlines and other sectors.
\end{abstract}


\begin{highlights}
\item Extracts themes and sentiment from tens of thousands of online reviews for 80 airlines
\item Identifies the evaluation indicators of Airline service quality ranking
\item Facilitates objective and clearly evaluation of Airline service quality ranking
\item Visualises Airline service quality ranking, and tests the sensitivity of the model
\end{highlights}


\maketitle

\section{Introduction}
\label{sec:section1}
An airline's service quality ranking significantly influences its profitability and sustainability in the intensely competitive aviation industry \citep{mousavi2020corporate}. Consumers demonstrate acute sensitivity towards aspects such as airline reputation \citep{PARK2004435}, loyalty programs \citep{jiang2016investigation}, safety standards \citep{baisya2004customer}, and punctuality \citep{elliott1993service}. It is imperative for airlines to comprehend customer expectations and deliver exemplary services \citep{duncan2017customer}. While numerous quality rankings derive from consumer surveys, the availability of such surveys is limited, and they often suffer from subjectivity, inconsistent standards, and challenging comparability.

To navigate the intricacies of decision-making in service quality ranking, scholars have increasingly adopted multi-criteria decision-making (MCDM) methodologies. MCDM, a subdivision of operations research, addresses intricate decisional challenges encompassing multiple, sometimes conflicting, criteria \citep{taherdoost2023multi}. MCDM facilitates the evaluation and comparison of alternatives based on their performance across various criteria, enabling the selection of the optimum or most favored option as per a defined preference model \citep{sahoo2023comprehensive}. Practical applications of MCDM span diverse domains, including occupational health risk evaluation \citep{thokala2016multiple}, supply chain network analysis \citep{gul2018review}, and decision-making in healthcare scenarios \citep{das2022building}. However, traditional MCDM methods grapple with challenges such as the inadequate amount of data and inefficiencies in ranking mechanisms, lack of ranking stability, and an over-reliance on historical quantitative data \citep{sotoudeh2022applications}.

In light of these challenges, there is an urgent necessity to augment traditional MCDM approaches by incorporating supplementary information sources. Text-mining techniques, in recent developments, have proven to be invaluable for gleaning insights from extensive textual datasets \citep{antons2020application,hassani2020text}. Specifically, in aviation service quality assessment and similar fields, researchers have harnessed text mining to delve into customer reviews, feedback, and sentiments \citep{lucini2020text,kumar2021applications}. This qualitative aspect, when amalgamated with MCDM, enriches the decision-making framework by capturing subtle details that may elude purely quantitative data analyses \citep{hashemkhani2020synergies}.

In the realm of service quality rankings, the integration of text-mining techniques with MCDM has yielded insightful studies. \cite{park2023combined} combined data envelopment analysis with text mining to analyze online text reviews and evaluate customer satisfaction for the top 20 global airlines in 2020. Similarly, \cite{singh2022does} combined text mining methods such as sentiment analysis and topic modeling to analyze the impact of service quality captured through customer reviews of logistics service providers on different operational, financial and aggregate performance. \cite{eshkevari2022end} integrated semantic mining and MCDM to design an end-to-end ranking method to rank the quality of hotel services, facilities, and amenities based on customer reviews.

The research aims to explore the crowd wisdom \citep{prelec2017solution} in tens of thousands of online reviews, provide airlines with insights into current business issues, and provide customers with a clear and comprehensive ranking list of airline service quality. More specifically, our contributions are as follows: (i) Using text mining and sentiment analysis techniques to extract tens of thousands of online airline service reviews in different languages, which may be the largest data set in history, and determine different evaluation criteria; (ii) Proposing a completely new The comprehensive MCDM model takes into account the interrelationship between standards, which can check airline quality in a timely and comprehensive manner and maximize ranking efficiency and stability; (iii) In the form of a directed topological hierarchical graph, 80 airlines around the world are displayed Visual ranking results of airline service quality.

The remainder of this paper is organized as follows. Section \ref{LR} provides a review of past research and progress. Section \ref{Methods} describes the text-mining and MCDM method. Section \ref{Results} presents the results of an empirical case study on airline service quality and discusses the outcomes. Section \ref{Discussion} discusses the obtained results and findings. Finally, Section \ref{Conclusion} concludes this paper by outlining the model's limitations and making recommendations for further research.

\section{Literature Review}
\label{LR}

Service quality encapsulates a customer's evaluation of a service provider's efficiency and the effectiveness of its offerings \citep{huang2010effect}, as well as the interaction between these aspects and other potential influencing variables \citep{gursoy2005us}. It could also be regarded as the customer's comprehensive appraisal of a service procedure \citep{chen2005examining}. In the airline industry, service quality is a critical determinant of customer satisfaction \citep{jiang2016investigation,chen2005examining,namukasa2013influence}. The initial step in creating and delivering high-caliber service is an in-depth understanding of individual customer expectations \citep{zeithaml1990delivering}. However, due to its intangible nature, variability, uniqueness, and the intricacies involved in consumer knowledge and experience, service quality is very difficult to characterize and quantify precisely \citep{laming2014customer}.

Recent research has segmented the criteria for evaluating airline service quality into various subcategories, encompassing aspects like pricing \citep{gourdin1988bringing}, safety \citep{gilbert2003passenger}, punctuality \citep{ostrowski1993service}, in-flight catering \citep{elliott1993service}, baggage handling \citep{truitt1994evaluating}, seating comfort \citep{bellizzi2020online}, pre-boarding services \citep{bellizzi2020online}, in-flight amenities \citep{bellizzi2020online}, and the effectiveness in addressing and resolving complaints \citep{liou2007non}. Historically, the assessment of service quality in airlines has been conducted through various methodologies. For instance, \cite{chen2008investigating} explored the nexus between airline passengers' satisfaction, perceived value, and service quality via structural equation modeling; \cite{leon2020fuzzy} utilized fuzzy partitioning to evaluate the technical and functional quality of the US airline sector in relation to passenger satisfaction; and \cite{hu2016quality} applied the Kano model to assess quality risks in Taiwanese airlines, noting that subpar service quality leads to customer dissatisfaction.

The SERVQUAL scale is the most comprehensive and widely employed model for understanding service quality \citep{parasuraman1985conceptual}. This model includes five topics: tangibility, reliability, responsiveness, assurance, and empathy, each serving to effectively gauge clients' perceptions of service quality. Tangibility refers to the physical presentation of services, such as in-flight technology and catering standards. Reliability pertains to the airline's credibility, encompassing aircraft safety and crew expertise. Responsiveness involves the interaction and communication between the crew (both on the ground and onboard) and passengers. Assurance is indicative of the certainty and confidence provided by the airline's services, including aspects like crew language proficiency. Lastly, empathy deals with the airline's approach to handling customer complaints and its ability to provide personalized service \citep{parasuraman1985conceptual,parasuraman1988servqual}. Understanding how customer expectations are shaped and their perception of service quality is fundamental to the SERVQUAL. A positive quality impression from the customer arises only when the service provider meets or surpasses these expectations \citep{robledo2001measuring}.

In the realm of airline service quality, the integration of text mining techniques with the established SERVQUAL scale has emerged as a formidable approach to delve into the nuances of service quality assessment. Text mining, a method that extracts pertinent information from unstructured text data such as online reviews, leverages natural language processing, machine learning, and statistical approaches \citep{jo2019text}. Latent semantic analysis (LSA) is a technique that analyzes the relationships between a set of documents and the terms they contain by producing a set of concepts related to the documents and terms \citep{muzumdar2024latent}. Employing Singular Value Decomposition (SVD), LSA adeptly reduces the complexity of the document-term matrix, whilst maintaining the intrinsic similarity structure among the documents and terms \citep{wagire2020analysis}. Its applicability spans information retrieval, document clustering, and topic modeling, with its efficacy substantiated in service quality evaluation \citep{vencovsky2020service,badanik2023sentimental,shah2021listening}. By applying text mining to customer reviews and feedback, researchers gain the ability to extract valuable insights and sentiments, providing a more nuanced understanding of customers' perceptions mapped to the structured topics of SERVQUAL \citep{mejia2021service,chatterjee2022measuring}. Automated extraction of SERVQUAL topics from online reviews enables comparison across various airlines, regions, and customer demographics \citep{bogicevic2017visual}. For instance, \cite{tian2020new} utilized social media data to assess airline service quality through SERVQUAL indicators, while \cite{sezgen2019voice} analyzed over 5,000 passenger reviews from TripAdvisor, correlating them with the SERVQUAL scale to facilitate customer satisfaction analysis for airline managers. This fusion of SERVQUAL and text mining represents a significant stride in enhancing our comprehension of airline service quality, fostering an evaluation that is both exhaustive and adaptive, resonating with the dynamic expectations of passengers in the ever-evolving aviation sector.

TOPSIS \citep{hwang1981methods,yoon1987reconciliation} is a comprehensive ranking method based on choosing the ideal solution that has the minimum geometric distance from the positive ideal solution and the maximum distance from the set of negative ideal solutions \citep{chen1992fuzzy}. The VIKOR technique resolves multi-objective decision-making situations that have conflicting criteria \citep{opricovic1998multicriteria,opricovic2004compromise}. Finding a compromise solution involves ranking and selecting options in the context of the competing criteria \citep{opricovic2004compromise}. AISM is an extension of the ISM method that breaks down a complex system's constituent parts into smaller parts, arranges those parts in a cause-and-effect hierarchy through a series of boolean and topological operations, and then identifies the topological structure hierarchical graph \citep{LI2023100110}. Based on the result-oriented hierarchical ranking rules of ISM, AISM adds a cause-oriented hierarchical ranking to provide a collection of directed topological diagrams that are in opposition to the ISM ranking rules \citep{biao2020extensibility}. By using AISM to analyze complicated systems, the structure of these systems can be determined without affecting their functionality, and a simple, hierarchical directed topology diagram can be produced.

Our methodology synergizes LSA, TOPSIS, VIKOR, and AISM to formulate an equitable and comprehensible quality rating system. LSA facilitates data collection analysis and subsequent sentiment assessment using a sentiment dictionary. The TOPSIS and VIKOR methods are employed for data reduction and to ascertain average distances to ideal, compromise, and extremal solutions. Additionally, a partial order operation on the data is executed to derive the corresponding relationship matrix. The culmination of this process, aided by the AISM approach, is a directed hierarchical topological plot, effectively constituting the airline service quality ranking, achieved through continuous clip-forcing approximation.

\section{Methods}
\label{Methods}

\begin{figure}
\centering
\includegraphics[width=0.8\linewidth]{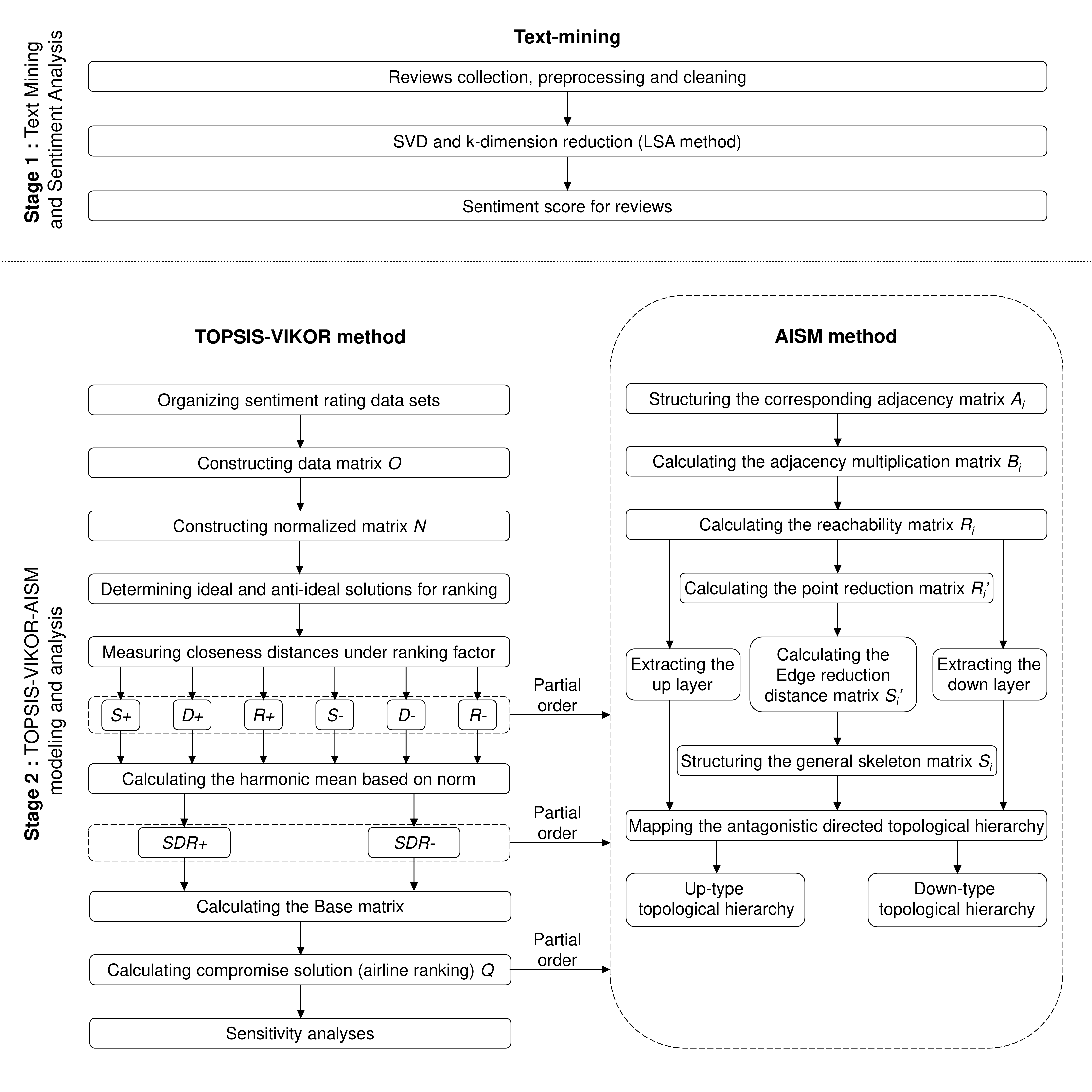}
\caption {Diagram of the text-mining method and TOPSIS-VIKOR-AISM model process.}
\label{fig:figure1}
\end{figure}

Figure \ref{fig:figure1} depicts the complete model development procedure.

\subsection{Data collection and cleaning}

Airline service rankings have long been of interest. Using data published by the international air transport ranking organization Skytrax, we selected 80 airlines for analysis. Details of these airlines are presented in Table \ref{tab:table4}.

\begin{table}
\centering
\caption{80 airlines analyzed in our study.}
\resizebox{\linewidth}{!}{
\begin{tabular}{lp{7cm}llp{7cm}}
\toprule
Number & \multicolumn{1}{c}{Airline} &  & Number & \multicolumn{1}{c}{Airline}   \\ \midrule
A1     & Aegean Airlines             &  & A41    & IndiGo                        \\
A2     & Aer Lingus                  &  & A42    & Japan Airlines                \\
A3     & Air Arabia                  &  & A43    & Jet2 Airlines                      \\
A4     & Air Astana                  &  & A44    & JetBlue Airways               \\
A5     & Air Canada                  &  & A45    & Jetstar Airways               \\
A6     & Air France                  &  & A46    & Jetstar Asia                  \\
A7     & Air New Zealand             &  & A47    & KLM Royal Dutch Airlines      \\
A8     & Air Transat                 &  & A48    & Korean Air                    \\
A9     & AirAsia                     &  & A49    & Kuwait Airways                \\
A10    & airBaltic                   &  & A50    & LATAM                         \\
A11    & Alaska Airlines             &  & A51    & LOT Polish                    \\
A12    & American Airlines           &  & A52    & Lufthansa                     \\
A13    & ANA All Nippon Airways      &  & A53    & Malaysia Airlines             \\
A14    & Asiana Airlines             &  & A54    & Oman Air                      \\
A15    & Austrian Airlines           &  & A55    & Peach                         \\
A16    & Azerbaijan Airlines         &  & A56    & Philippine Airlines           \\
A17    & Azul Airlines               &  & A57    & Qantas Airways                \\
A18    & Bangkok Airways             &  & A58    & Qatar Airways                 \\
A19    & British Airways             &  & A59    & Rex Airlines                  \\
A20    & Brussels Airlines           &  & A60    & Royal Air Maroc               \\
A21    & Cathay Pacific Airways      &  & A61    & Royal Brunei Airlines         \\
A22    & China Airlines              &  & A62    & Ryanair                       \\
A23    & China Southern Airlines     &  & A63    & SAS Scandinavian              \\
A24    & Delta Air Lines             &  & A64    & Saudi Arabian Airlines        \\
A25    & EasyJet                     &  & A65    & Scoot                         \\
A26    & Emirates                    &  & A66    & Singapore Airlines            \\
A27    & Ethiopian Airlines          &  & A67    & South African Airways         \\
A28    & Etihad Airways              &  & A68    & Southwest Airlines            \\
A29    & Eurowings                   &  & A69    & SunExpress                    \\
A30    & EVA Air                     &  & A70    & Swiss International Air Lines \\
A31    & Fiji Airways                &  & A71    & TAP Portugal                  \\
A32    & Finnair                     &  & A72    & Thai Airways                  \\
A33    & flyDubai                    &  & A73    & Turkish Airlines              \\
A34    & Flynas                      &  & A74    & United Airlines               \\
A35    & Garuda Indonesia            &  & A75    & Vietnam Airlines              \\
A36    & Gulf Air                    &  & A76    & Virgin Atlantic               \\
A37    & Hainan Airlines             &  & A77    & Virgin Australia              \\
A38    & Hawaiian Airlines           &  & A78    & Vistara                       \\
A39    & Hong Kong Airlines          &  & A79    & Vueling Airlines              \\
A40    & Iberia                      &  & A80    & WestJet                       \\ \bottomrule
\end{tabular}}
\label{tab:table4}
\end{table}

We embarked on a comprehensive collection of online reviews for these 80 airlines, sourcing from various travel websites in multiple languages, such as TripAdvisor ({\href{tripadvisor.com}{tripadvisor.com}}), Skytrax ({\href{skytraxratings.com}{skytraxratings.com}}), Trip ({\href{Trip.com}{Trip.com}}), and Qunar ({\href{qunar.com}{qunar.com}}), as well as from prominent social media platforms like X ({\href{x.com}{x.com}}), Facebook ({\href{facebook.com}{facebook.com}}), Weibo ({\href{weibo.com}{weibo.com}}), and Xiaohongshu ({\href{xiaohongshu.com}{xiaohongshu.com}}). These reviews encompassed a diverse range of service aspects, including but not limited to, flight experience, staff behavior, food quality, and seat comfort. A minimum of 200 reviews were gathered for each airline, amassing a substantial dataset of 61,087 reviews. The data collection strategy was meticulously designed to amass a robust and representative dataset, reflecting the breadth of customer opinions and preferences concerning airline service quality. We employed web scraping tools like BeautifulSoup \citep{richardson2007beautiful} in Python and Selenium, to methodically extract reviews from these websites. Additionally, we utilized the Google Translate API ({\href{translate.google.com/}{translate.google.com/}}) to translate reviews from various languages, including Chinese, French, Spanish, Arabic and etc., into English.

In the subsequent phase, we undertook the preprocessing of text data, aiming to eliminate noise and superfluous elements such as punctuation, stopwords, numbers, URLs, and more. This process also involved the application of stemming and lemmatization techniques, streamlining words to their base forms and harmonizing the vocabulary. The overarching goal of this data cleaning procedure was to enhance the textual data's quality and coherence, while simultaneously reducing the dimensionality of the term-document matrix. For this task, we utilized the NLTK package \citep{loper2002nltk} in Python. Furthermore, we meticulously conducted manual reviews to identify and rectify any spelling and grammatical inaccuracies within the reviews, ensuring the veracity and precision of the text data.

To demonstrate the efficacy of our data cleaning process, we present exemplars of both the original and cleaned reviews in Table \ref{tab:table0}. As illustrated, the data cleaning operation effectively expunged extraneous symbols and words, as well as streamlined sentences. This crucial step substantially augmented the efficiency of the subsequent analytical processes.

\begin{table}[pos=H]
\centering
\caption{Examples of the original reviews and the cleaned reviews.}
\resizebox{\linewidth}{!}{
\begin{tabular}{cp{9cm}p{9cm}}
\toprule
Number & Original Review                                                                                                                                                                                                                                      & Cleaned Review                                                                                                                                    \\ \midrule
1      & I flew with Airline A from   London to New York and I had a great experience. The arrival service were helpful, the food was delicious, and the seat was comfortable. I would definitely fly with them again.                                 & fly Airline A London New York great experience arrival service  helpful food delicious seat comfortable definitely fly again                                              \\
2      & Airline B is the worst airline  ever. They cancelled my flight without any notice and refused to refund me. They have no customer service and no respect for their passengers. I will never use them again.                                     & Airline B worst airline ever cancel flight without notice refuse refund no customer service no respect passenger never use                                       \\
3      & Airline C has excellent service   and value for money. The flights are always on time and the cabin crew are very professional and courteous. The entertainment system is also very good  and has a lot of options. I highly recommend Airline C! :)

& Airline C excellent service value money flights on time cabin crew very professional courteous entertainment system very good lot option highly recommend Airline C\\ \bottomrule
\end{tabular}}
\label{tab:table0}
\end{table}

\subsection{Identification and quantification of evaluation indicators}

Employing the framework of the five SERVQUAL topics and incorporating additional research and practical insights, we devised fifteen service quality assessment criteria across five categories, as delineated in Table \ref{tab:table2}.

\begin{table}[pos=H]
\centering
\caption{Criteria are selected in the study.}
\resizebox{0.5\linewidth}{!}{
\begin{tabular}{ccl}
\toprule
Topic & Indicator & Evaluation Indicators \\ \midrule
\multirow{3}{*}{Tangibility} & $C_1$ & Seat comfort \\ 
& $C_2$ & Quality of meals \\ 
& $C_3$ & In-flight entertainment \\ 
 \midrule
\multirow{3}{*}{Reliability} & $C_4$ & Staff professionalism \\ 
& $C_5$ & Aircraft punctuality \\ 
& $C_6$ & Aircraft safety \\ \midrule
\multirow{3}{*}{Responsiveness} & $C_7$ & Staff attitudes \\ 
& $C_8$ & Service attentiveness / efficiency \\ 
& $C_9$ & Problem-solving skills \\ \midrule
\multirow{3}{*}{Assurance} & $C_{10}$ & Staff skills \\ 
& $C_{11}$ & Staff grooming \\ 
& $C_{12}$ & Service fairness \\ \midrule
\multirow{3}{*}{Empathy} & $C_{13}$ & Vulnerable group services \\ 
& $C_{14}$ & Pre-boarding procedures \\ 
& $C_{15}$ & Transfer and arrival services \\ \bottomrule
\end{tabular}
}
\label{tab:table2}
\end{table}

\begin{figure}
\centering
\includegraphics[width=0.5\linewidth]{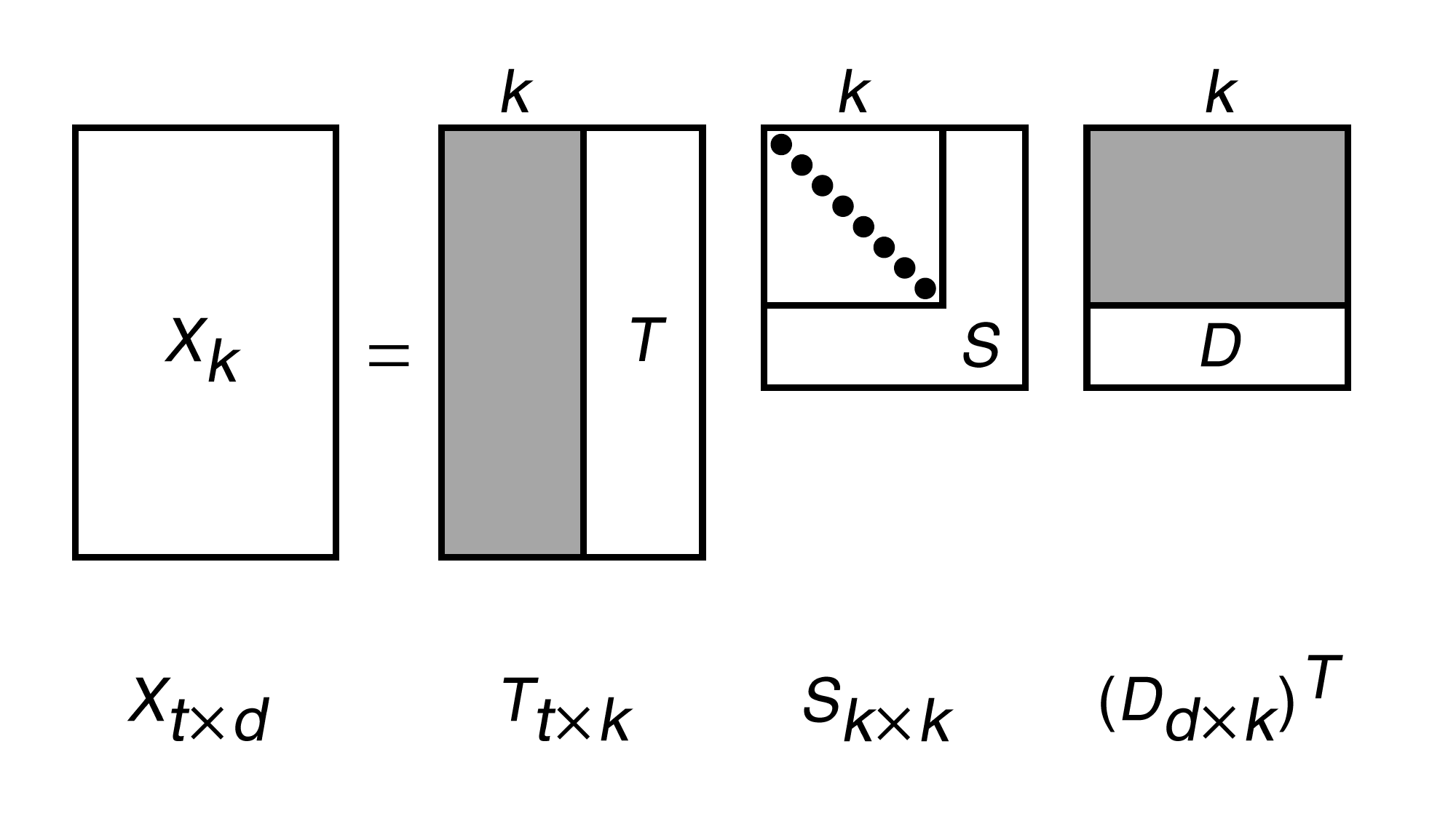}
\caption {The process of decomposition and truncation (adopted from \cite{Berry1995573}).}
\label{fig:figure0}
\end{figure}

Our text-mining endeavor was directed towards discerning pivotal themes and affective responses from customer feedback, quantifying these through numerical assessments. Utilizing the LSA technique, we extracted salient topics and sentiments from the amassed reviews. LSA, a method employing SVD, serves to reduce the dimensionality of a term-document matrix, thereby unveiling the underlying semantic structure of the text. This analysis comprises three primary stages. Initially, it involves representing documents and terms within a vector space, creating a term-document matrix that captures the concurrent occurrence of terms within documents. This matrix is then subject to term frequency-inverse document frequency (tf-idf) weighting, accentuating terms that distinctively describe the document's content. The final stage involves SVD, which decomposes the term-document matrix, denoted as $X_{t \times d}$, into three matrices: $T_{t \times m}$ (a column-orthogonal matrix with topics represented by $m$), $S_{m \times m}$ (a diagonal matrix containing singular values in descending order), and $D_{d \times m}$ (a transposed column-orthogonal matrix). Here, $t$ and $d$ represent the number of terms and documents, respectively. These matrices are truncated to a chosen number of topics, $k$, to filter out noise and distill latent semantic linkages within the text corpus. The decomposition and truncation process is depicted in Fig. \ref{fig:figure0}.

In constructing the term-document matrix from the curated reviews, each row corresponded to a word, and each column to a review. We employed the Scikit-learn package \citep{kramer2016scikit} in Python for the LSA procedure, selecting the top 'k' topics based on singular values and deriving the pertinent topic keywords and scores for each review.

For sentiment analysis, we utilized a scoring dictionary that allocated a value to words based on their polarity and intensity. The SentiWordNet 3.0 dictionary \citep{baccianella2010sentiwordnet}, a lexical resource tailored for opinion mining, was employed. It assigns objective scores to each synset (a set of synonyms) in WordNet. Words in the reviews were aligned with synsets in the dictionary to obtain word-specific sentiment scores. We then averaged these scores for each review, calculating an aggregate sentiment score. Additionally, the TextBlob package \citep{loria2018textblob} in Python facilitated sentiment analysis, assigning a polarity score from 0 (negative) to 5 (positive) to each review. Table \ref{tab:table000} elucidates the correlation between sample emotion words and their respective scores.

\begin{table}[pos=H]
\centering
\caption{Basic correspondence between each sample emotion word and emotion score.}
\begin{tabular}{ll}
\toprule
Emotion word                & Score \\ \midrule
excellent/amazing/wonderful & 5     \\
good/nice/happy             & 4     \\
fair/okay/pleasant          & 3     \\
average/neutral/indifferent & 2     \\
bad/awful/terrible          & 1     \\
worst/horrible/disgusting   & 0     \\ \bottomrule
\end{tabular}
\label{tab:table000}
\end{table}

A review's score, situated between two sentiment thresholds, indicates the presence of both positive and negative elements or words with varying intensity levels. For instance, a score of 4.3, hovering between 4 and 5, suggests a predominantly positive review but not exceedingly so. Such a review might feature highly positive words like "excellent" or "amazing", coupled with moderately positive terms like "good" or "nice". Alternatively, it might include positive words alongside neutral or marginally negative terms, such as "okay" or "average".

To exemplify our text-mining methodology, Table \ref{tab:table00000} presents sample extracts from the reviews. As shown in the table, the reviews were labeled according to the five SERVQUAL topics and broken down into specific sub-indicators.  Sentiments are classified as positive, negative, or neutral, based on their polarity scores. For example, the review "Airline A has excellent service and value for money" received a sentiment score of 3.75, indicative of a positive sentiment.

\begin{table}[pos=H]
\centering
\caption{Examples of topics and sentiments extracted from the reviews.}
\resizebox{\linewidth}{!}{
\begin{tabular}{cp{8cm}lllll}
\toprule
\multicolumn{1}{l}{Number} & Review                                                                                                              & Topic          & Indicator & Indicator Score & Sentiment & Polarity Score \\ \midrule
1                          & Airline A has excellent service and value for money                                                               & Assurance   &  Staff skills  & 4         & Positive  & 4.5           \\
2                          & Airline B is the worst airline ever. They cancelled my flight without any notice and refused to refund me         & Reliability   & Aircraft punctuality & 0.5        & Negative  & 1           \\
3                          & Airline C has a very good  entertainment system and a lot of options                                               & Tangibles   & In-flight entertainment   & 4         & Positive  & 4            \\
4                          & I flew with Airline A from London to New York and I had a great experience. The arrival service were helpful & Empathy  & Transfer and arrival services      & 4.5        & Positive  & 4           \\
5                          & Airline B has no customer ervice and no respect for their passengers                                             & Responsiveness & Staff attitudes &  1        & Negative  & 1.5           \\
6                          & Airline C's seat was uncomfortable and the food was terrible                                                      & Tangibles  & Seat comfort  & 1.5        & Negative  & 1           \\ \bottomrule
\end{tabular}}
\label{tab:table00000}
\end{table}

We created an original evaluation matrix $O$ based on $m$ evaluation objects and $n$ evaluation indicators, where $x_{ij}$ is the average score of each evaluation object in the evaluation indicator. An example is presented in Table \ref{tab:table3}.

In this matrix, $n$ and $m$ are the number of airlines and indicators, respectively. Thus, 
\begin{equation}
O=x_{i j},\quad i=1,\cdots,n;\quad j=1,\cdots,m.
\label{eq:equation1}
\end{equation}

\begin{table}[pos=H]
\centering
\caption{Original assessment matrix.}
\begin{tabular}{ccccc} 
\toprule 
Airline & $C_1$ & $C_2$ & $\cdots$ & $C_{m}$ \\ \midrule
$A_1$ & $x_{11}$ & $x_{12}$ & $\cdots$ & $x_{1m}$ \\ 
$A_2$ & $x_{21}$ & $x_{22}$ & $\cdots$ & $x_{2m}$ \\ 
$\cdots$ & &$\cdots$ & & \\ 
$A_{n}$ & $x_{n1}$ & $x_{n2}$ & $\cdots$ & $x_{nm}$ \\
\bottomrule
\end{tabular}

\label{tab:table3}
\end{table}

\subsection{Construction of TOPSIS-VIKOR-AISM model}

To produce $N$, the initial evaluation matrix $O$ is normalized as follows:
\begin{equation}
N=r_{i j}=\frac{x_{i j}}{\sqrt{\sum_{k=1}^m x_{kj}^2 }},\quad i=1,\cdots,n, j=1,\cdots,m.
\label{eq:equation2}
\end{equation}

We normalize the compiled data on airline service quality and arrange it into a raw matrix covering the different airlines, indications, and scores. We compute utility values $S^+$, $S^-$, Euclidean distances $D^+$, $D^-$, and regret values $R^+$, $R^-$ using the TOPSIS and VIKOR. We then compute the mean of the distances to positive and negative ideal points, $SDR^+$ and $SDR^-$, from the compromise solution $Q$. Three sets of adversarial hierarchical topologies are successively computed for the topologies acquired using the AISM approach, as depicted on the right side of Figure \ref{fig:figure1}. An intuitive hierarchical link between airline service quality and $S^+$, $S^-$, $D^+$, $D^-$, $R^+$, $R^-$, $SDR^+$, $SDR^-$, and $Q$ is obtained.

\textbf{Step 1:} Calculate the positive and negative ideal solutions $F_j^+$ and $F_j^-$.
\begin{equation}
\begin{aligned}
    F_j^+=&[t_1^+,\cdots,t_j^+,\cdots,t_m^+]=[\text{max}{(t_{11},t_{21},\cdots,t_{n1})},\cdots,\\
&\text{max}{(t_{1j},t_{2j},\cdots,t_{nj})},\cdots, \text{max}{(t_{1m},t_{2m},\cdots,t_{nm})}];
\end{aligned}
\label{eq:equation7}
\end{equation}
\begin{equation}
\begin{aligned}
F_j^-=&[t_1^-,\cdots,t_j^-,\cdots,t_m^-]=[\text{min}{(t_{11},t_{21},\cdots,t_{n1})},\cdots,\\
&\text{min}{(t_{1j},t_{2j},\cdots,t_{nj})},\cdots,\text{min}{(t_{1m},t_{2m},\cdots,t_{nm})}].
\end{aligned}
\label{eq:equation8}
\end{equation}

\textbf{Step 2:} Calculate the alternative solution and compute the positive and negative ideal solution distances $D^+$ and $D^-$.
\begin{equation}
\begin{aligned}
D^+=\sqrt{\sum_{j=1}^m (t_{ij} - t_j^+)^2},\quad i=1,\cdots,n;
\label{eq:equation9}
\end{aligned} 
\end{equation}
\begin{equation}
\begin{aligned}
D^-=\sqrt{\sum_{j=1}^m (t_{ij} - t_j^-)^2},\quad i=1,\cdots,n.
\label{eq:equation10}
\end{aligned} 
\end{equation}

\textbf{Step 3:} Determine the best and worst evaluation functions $f_j^+$ and $f_j^{-}$.
\begin{equation}
f_j^+=\max _i t_{i j},\quad i=1,\cdots,n, j=1,\cdots,m;
\label{eq:equation11}
\end{equation}
\begin{equation}
f_j^{-}=\min _i t_{i j},\quad i=1,\cdots,n, j=1,\cdots,m.
\label{eq:equation12}
\end{equation}

\textbf{Step 4:} Calculate the weighted normalized Manhattan distance $S_i$ and the weighted normalized Chebyshev distance $R_i$.
\begin{equation}
S_i=\sqrt{\sum_{j=1}^m w_j\frac{f_j^+-f_{ij}}{f_j^+-f_j^-}},\quad i=1,\cdots,n, j=1,\cdots,m;
\label{eq:equation13}
\end{equation}
\begin{equation}
R_i=\max_j[\frac{w_j(f_j^+-f_{ij})}{(f_j^+-f_j^-)}],\quad i=1,\cdots,n, j=1,\cdots,m,
\label{eq:equation14}
\end{equation}
where, $w_j$ is the weight of each indicator, indicating its relative importance. 

\textbf{Step 5:} Calculate the compromise solution. If we suppose the following Equation (\ref{eq:equation15}) holds.
\begin{equation}
S^-=\min_i S_i;~S^+=\max_i S_i;~R^-=\min_i R_i;~R^+=\max_i R_i.
\label{eq:equation15}
\end{equation}

Then we define that $SDR^+$ and $SDR^-$ are the harmonic means of the distances of each element to the positive and negative ideal points respectively. So the compromise solution $Q_i$ can be determined as follows using Equation (\ref{eq:equation16}).
\begin{equation}
\begin{aligned}
Q_i=&(1-k)\left(\frac{S D R_i^{+}-\text{min}\left(S D R_i^{+}\right)}{\text{max}\left(S D R_i^{+}\right)-\text{min}\left(S D R_i^{+}\right)}\right)\\
&+k\left(\frac{\text{max}\left(S D R_i^{-}\right)-S D R_i^{-}}{\text{max}\left(S D R_i^{-}\right)-\text{min}\left(S D R_i^{-}\right)}\right).
\end{aligned}
\label{eq:equation16}
\end{equation}

The allocation coefficient $k$ is typically set to 0.5, representing the weight of the maximum group utility strategy being adopted.

\textbf{Step 6:} Build the adjacency relationship matrix. The partial order relationship operation transfers the evaluation matrix on the left to the relationship matrix on the right based on the partial order rules. These are the guidelines for partial orders. For an evaluation matrix $D$ containing $m$ columns, all columns, i.e., the indicator dimensions, have the same properties and are thus comparable. To compare the advantages and disadvantages of dimensions, we must determine whether the indicators are positive or negative. When larger values are better and smaller values are worse, the indicator is said to be positive. Such indicators are denoted as $p_1, p_2, \ldots, p_m$. When larger values are worse and smaller values are preferable, the indicator is said to be negative, denoted as $q_1, q_2, \ldots, q_m$. For any two rows $x$, $y$ in $D$, if the indicators are positive, we have
\begin{equation}
d_{(x, p_1)} \ge d_{(y, p_1)} , d_{(x, p_2)} \ge d_{(y, p_2)},~\cdots,~ d_{(x, p_m)} \ge d_{(y, p_m)}.
\label{eq:equation17}
\end{equation}

The partial order connection between $x$ and $y$ is noted as $x \prec y$, which suggests that $y$ is superior to $x$ if the aforementioned rule is satisfied. The internal relationship between the influencing factors can then be discovered by expert scoring in accordance with the various evaluation indices. The element $a_{ij}$ in the adjacency matrix $A$ can then be written as
\begin{equation}
a_{ij}=\left\{\begin{array}{ll}
0 ,& x\prec y; \\
1 ,& \text {No perfect relationship between } x \text { and } y \text {. }
\end{array}\right\}
\label{eq:equation18}
\end{equation}

\textbf{Step 7:} The accessibility matrix reflects the lengths of the paths connecting any two nodes in a directed connected network. It is necessary to first compute the multiplicative adjacency matrix $B$ for any given basic matrix.
\begin{equation}
B=A+I,
\label{eq:equation19}
\end{equation}
where $B$ is the multiplicative adjacency matrix and $I$ is the unit matrix. The concatenation of $B$ results in the reachable matrix $R$.
\begin{equation}
{B}^{{k}-1} \neq {B}^{{k}}={B}^{{k}+1}={R}.
\label{eq:equation20}
\end{equation}

The cycle in the reachable matrix $R$ is treated as a point in the point reduction operation on the reachable matrix $R$, from which a new reachable matrix $R'$ can be generated. The edge reduction technique is applied to $R'$ to obtain the skeleton matrix $S'$, which essentially eliminates redundant routes.
\begin{equation}
S^{\prime}=R^{\prime}-\left(R^{\prime}-I\right)^2-I.
\label{eq:equation21}
\end{equation}

The general skeleton matrix $S$ can be obtained if the cyclic loops in $S^{\prime}$ are modeled as minimum daisy chains.

\textbf{Step 8:} Extraction of the hierarchy. For any accessibility matrix, there exist a reachable set $R$, a prior set $Q$, and a common set $T$, where $T=R\cap Q$. Consider a relational matrix $A$ as an illustration. For each element $e_i$ in $A$, there are two possible scenarios.

\begin{itemize}
\item All elements whose corresponding row value is 1 constitute the reachable set, and $R=(e_i)$;
\item The entire set of elements with a column value of 1 constitutes the prior set $Q(e_i)$. The common set of the reachable set $R=(e_i)\cap Q(e_i)$ and the prior set is called $T(e_i)$.
\end{itemize}

The layer extraction method is as follows.

\begin{itemize}
\item Extraction of the topology of up-type structures. The term 'result-first hierarchical extraction' refers to up-type structures, which follow the rule $T(e_i)=R(e_i)$. The main idea is to take the system components that make up the end result, place them in the top level, and then extract them via analogy;

\item Topology extraction of the down-type structures. We consider this to be a cause-based hierarchical extraction technique for structures which follow the rule $T(e_i)=Q(e_i)$. The fundamental idea is that the system components that are the root causes are first isolated and positioned at the bottom of the hierarchy, after which they are extracted via analogy.
\end{itemize}

\section{Results}
\label{Results}

\subsection{Analysis results of the proposed method}

\begin{figure}[pos=H]
\centering
\includegraphics[width=0.7\linewidth]{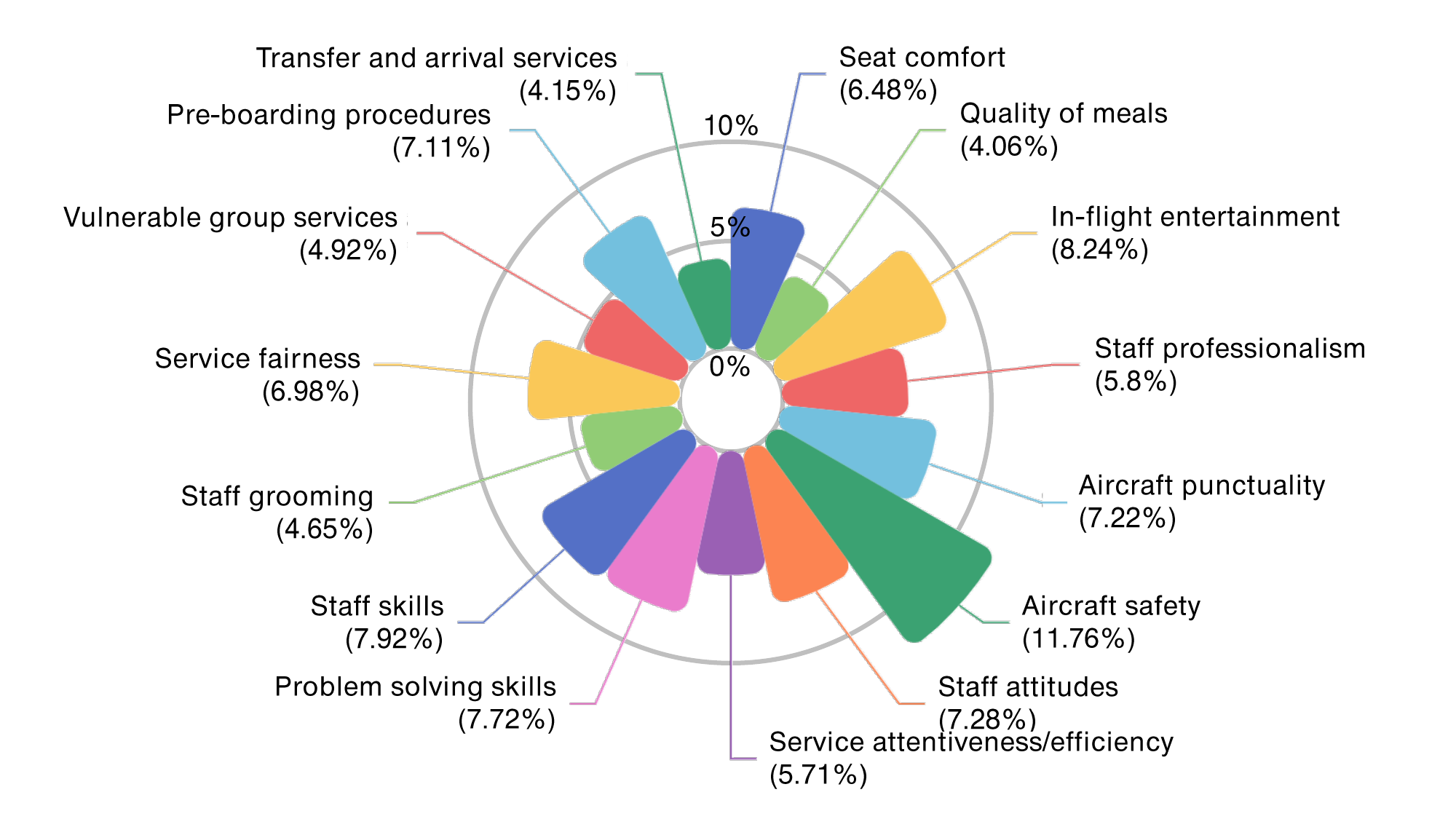}
\caption {The quantity distribution of each indicator.}
\label{fig:figure000}
\end{figure}

Following text mining, the distribution of the number of all online reviews mapped to each indicator can be found in Figure \ref{fig:figure000}. Notably, 'aircraft safety' emerges as the predominant keyword, constituting 11.78\% of the mentions. Other significant keywords include 'in-flight entertainment' (8.24\%) and 'staff skills' (7.92\%), highlighting focal areas of passenger concern.

The process of text mining has elucidated the top 20 keywords recurrent in customer feedback, as tabulated in Table \ref{tab:table010}. This table offers insightful patterns regarding customer perceptions and preferences in airline service quality. For instance, 'flight' surfaces as the most frequently cited keyword, with 22,340 occurrences, predominantly under the 'reliability' theme. This underscores a prevalent customer focus on flight performance aspects such as punctuality and safety. Similarly, the keyword 'service', appearing 20,980 times and classified under the 'assurance' theme, reflects the high value passengers place on aspects like staff proficiency and presentation, often manifesting in either commendations or criticisms of service quality.

\begin{table}
\centering
\caption{Top 20 most frequent keywords during text mining.}
\begin{tabular}{lll}
\toprule
Keyword       & Number & SERVQUAL Topic \\ \midrule
flight        & 22,340 & Reliability    \\
service       & 20,980 & Assurance      \\
staff         & 19,872 & Empathy        \\
seat          & 18,761 & Tangibles      \\
food          & 17,655 & Tangibles      \\
time          & 16,544 & Reliability    \\
entertainment & 15,435 & Tangibles      \\
customer      & 14,329 & Responsiveness \\
experience    & 13,211 & Assurance      \\
value         & 12,100 & Assurance      \\
refund        & 11,992 & Responsiveness \\
delay         & 11,881 & Reliability    \\
comfort       & 11,777 & Tangibles      \\
friendly      & 11,663 & Empathy        \\
helpful       & 11,632 & Empathy        \\
cancel        & 11,440 & Reliability    \\
respect       & 11,339 & Responsiveness \\
professional  & 11,221 & Assurance      \\
option        & 11,110 & Tangibles      \\
money         & 9,927  & Assurance      \\ \bottomrule    
\end{tabular}
\label{tab:table010}
\end{table}

Sentiment analysis yields the service satisfaction scores across five SERVQUAL thematic areas for 80 airlines, as detailed in Table \ref{tab:table0101}.

Equations (\ref{eq:equation9}, \ref{eq:equation10}, \ref{eq:equation15}) facilitate the calculation of positive and negative utility values ($S^+$, $S^-$), ideal solutions ($D^+$, $D^-$), and regret values ($R^+$, $R^-$). The harmonic mean distances from each element to the positive and negative ideal points, alongside the compromise solution ($Q_i$), are computed via Equation (\ref{eq:equation16}). These findings are enumerated in Table \ref{tab:table1}, with the far-right column presenting the rankings based on $Q_i$.

\begin{figure}[pos=H]
\centering
\includegraphics[width=0.5\linewidth]{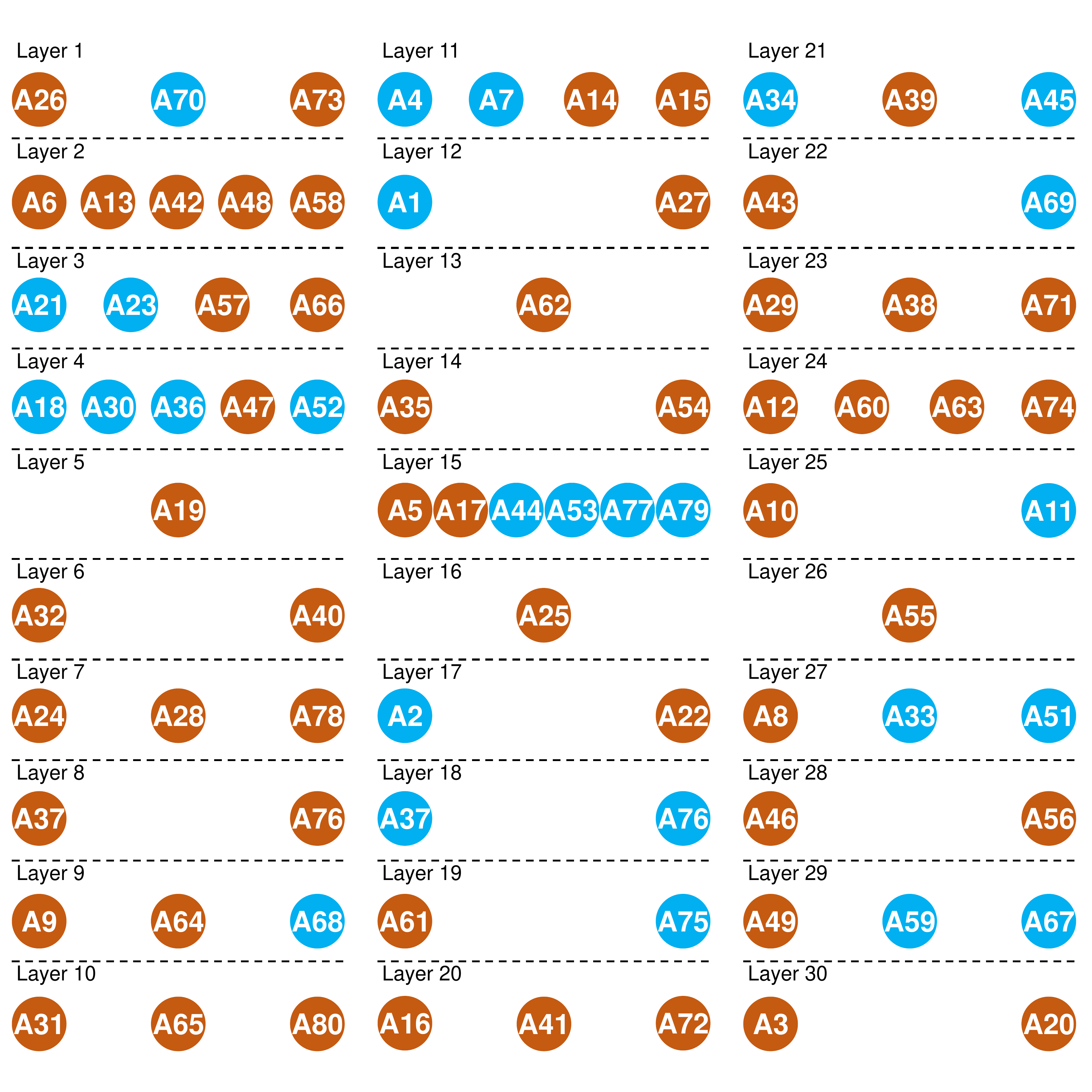}
\caption {Up-type directed topological hierarchy diagrams.}
\label{fig:figure2}
\end{figure}

\begin{figure}[pos=H]
\centering
\includegraphics[width=0.5\linewidth]{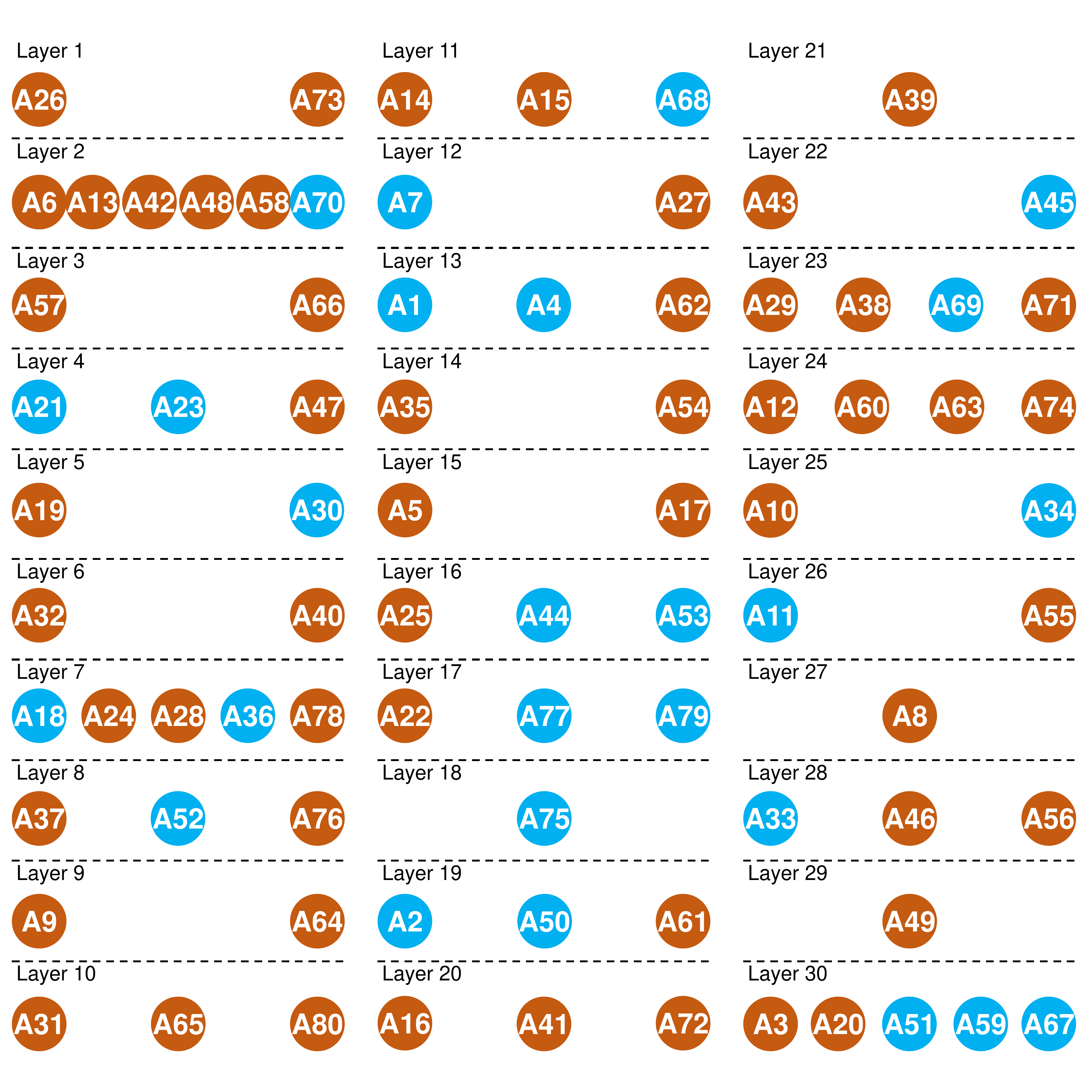}
\caption {Down-type directed topological hierarchy diagrams.}
\label{fig:figure3}
\end{figure}

Combining data for $D^{+}$, $D^{-}$, $S^{+}$, $S^{-}$, $R^{+}$, $R^{-}$, and subsequent pinch-force computations, the AISM analysis yields topological hierarchy diagrams, as depicted in Figures \ref{fig:figure2} and \ref{fig:figure3}. In these diagrams, brown circles symbolize airlines with stable rankings, while blue circles indicate those susceptible to fluctuating ratings. This analysis transcends mere ranking of airlines' service quality; it segments them into 30 distinct tiers, wherein airlines may share rankings. For example, Emirates (A26) and Turkish Airlines (A73), situated in the first tier, display comparable competencies in Tangibility and Empathy, albeit with minor differences in Reliability and Responsiveness Assurance. Such nuances in service quality competitiveness justify their placement within the same tier as per the AISM results. The final ranking, adhering to these principles, is exhibited in Table \ref{tab:table1010}.

\begin{table}[pos=H]
\centering
\caption{Service satisfaction scores across five thematic topics for 80 airlines.}
\resizebox{\linewidth}{!}{
\begin{tabular}{p{3cm}p{6cm}p{2.5cm}p{2.5cm}p{2.5cm}p{2.5cm}p{2.5cm}}
\toprule
Number & Airline                       & Tangibility & Reliability & Responsiveness & Assurance & Empathy \\ \midrule
A1     & Aegean Airlines               & 3.228       & 3.498       & 3.414          & 3.464     & 3.295   \\
A2     & Aer Lingus                    & 2.671       & 2.451       & 2.536          & 2.992     & 2.148   \\
A3     & Air Arabia                    & 1.101       & 1.152       & 1.186          & 1.574     & 1.169   \\
A4     & Air Astana                    & 3.380       & 3.312       & 3.481          & 3.599     & 2.924   \\
A5     & Air Canada                    & 2.468       & 2.502       & 2.975          & 2.975     & 2.857   \\
A6     & Air France                    & 4.730       & 4.561       & 4.730          & 4.932     & 4.392   \\
A7     & Air New Zealand               & 3.515       & 3.633       & 3.380          & 3.127     & 3.363   \\
A8     & Air Transat                   & 1.540       & 1.253       & 1.405          & 1.422     & 1.658   \\
A9     & AirAsia                       & 4.004       & 3.954       & 3.515          & 3.160     & 3.481   \\
A10    & airBaltic                     & 2.013       & 1.861       & 1.523          & 2.063     & 1.743   \\
A11    & Alaska Airlines               & 1.624       & 2.097       & 1.911          & 1.591     & 2.181   \\
A12    & American Airlines             & 2.030       & 1.895       & 1.945          & 1.844     & 1.523   \\
A13    & ANA All Nippon Airways        & 4.392       & 4.612       & 4.865          & 4.848     & 4.713   \\
A14    & Asiana Airlines               & 3.549       & 3.819       & 3.093          & 3.852     & 3.430   \\
A15    & Austrian Airlines             & 3.684       & 4.038       & 3.430          & 3.194     & 3.262   \\
A16    & Azerbaijan Airlines           & 2.367       & 2.637       & 2.435          & 2.485     & 2.435   \\
A17    & Azul Airlines                 & 2.823       & 2.755       & 2.586          & 2.722     & 2.907   \\
A18    & Bangkok Airways               & 4.089       & 3.903       & 3.852          & 3.954     & 4.426   \\
A19    & British Airways               & 4.291       & 4.494       & 4.089          & 4.257     & 4.527   \\
A20    & Brussels Airlines             & 1.135       & 1.270       & 1.338          & 1.473     & 1.456   \\
A21    & Cathay Pacific Airways        & 4.612       & 3.700       & 4.409          & 4.578     & 4.072   \\
A22    & China Airlines                & 2.907       & 2.570       & 2.654          & 2.705     & 2.789   \\
A23    & China Southern Airlines       & 4.021       & 3.937       & 4.376          & 4.544     & 4.544   \\
A24    & Delta Air Lines               & 4.426       & 3.954       & 4.443          & 4.122     & 4.072   \\
A25    & EasyJet                       & 3.059       & 2.435       & 2.502          & 2.907     & 2.890   \\
A26    & Emirates                      & 4.899       & 4.511       & 4.713          & 4.781     & 4.865   \\
A27    & Ethiopian Airlines            & 3.582       & 3.312       & 3.717          & 3.717     & 2.958   \\
A28    & Etihad Airways                & 4.308       & 3.987       & 3.768          & 4.257     & 4.257   \\
A29    & Eurowings                     & 2.316       & 2.013       & 2.097          & 1.709     & 1.759   \\
A30    & EVA Air                       & 4.207       & 3.937       & 4.139          & 4.392     & 4.376   \\
A31    & Fiji Airways                  & 3.262       & 3.329       & 3.498          & 3.650     & 3.768   \\
A32    & Finnair                       & 4.342       & 4.359       & 4.122          & 3.970     & 4.105   \\
A33    & flyDubai                      & 1.574       & 1.371       & 1.169          & 1.084     & 1.641   \\
A34    & Flynas                        & 2.350       & 2.232       & 2.030          & 2.165     & 2.030   \\
A35    & Garuda Indonesia              & 3.177       & 3.177       & 2.924          & 3.397     & 3.599   \\
A36    & Gulf Air                      & 3.954       & 4.038       & 4.156          & 3.768     & 4.274   \\
A37    & Hainan Airlines               & 4.021       & 4.207       & 3.464          & 3.920     & 4.409   \\
A38    & Hawaiian Airlines             & 1.641       & 2.013       & 2.300          & 2.283     & 1.962   \\
A39    & Hong Kong Airlines            & 1.861       & 1.878       & 2.249          & 2.046     & 2.620   \\
A40    & Iberia                        & 4.038       & 4.241       & 4.105          & 4.460     & 4.224   \\
A41    & IndiGo                        & 2.620       & 2.586       & 2.468          & 2.570     & 2.232   \\
A42    & Japan Airlines                & 4.426       & 4.814       & 4.848          & 4.899     & 4.578   \\
A43    & Jet2.com                      & 2.316       & 1.945       & 1.979          & 1.979     & 2.114   \\
A44    & JetBlue Airways               & 2.620       & 2.755       & 3.076          & 2.519     & 3.346   \\
A45    & Jetstar Airways               & 1.591       & 2.418       & 2.097          & 2.266     & 2.181   \\
A46    & Jetstar Asia                  & 1.658       & 1.236       & 1.439          & 1.405     & 1.422   \\
A47    & KLM Royal Dutch Airlines      & 4.376       & 4.426       & 4.409          & 4.156     & 4.443   \\
A48    & Korean Air                    & 4.797       & 4.747       & 4.764          & 4.696     & 4.730   \\
A49    & Kuwait Airways                & 1.506       & 1.405       & 1.135          & 1.371     & 1.253   \\
A50    & LATAM                         & 2.300       & 2.924       & 2.840          & 2.401     & 2.570   \\
A51    & LOT Polish                    & 1.557       & 1.338       & 1.506          & 1.321     & 1.574   \\
A52    & Lufthansa                     & 3.633       & 3.532       & 4.004          & 4.308     & 4.426   \\
A53    & Malaysia Airlines             & 3.025       & 2.367       & 2.890          & 2.705     & 2.705   \\
A54    & Oman Air                      & 3.262       & 3.380       & 3.498          & 2.924     & 3.262   \\
A55    & Peach                         & 1.692       & 1.945       & 1.962          & 1.743     & 1.844   \\
A56    & Philippine Airlines           & 1.405       & 1.304       & 1.658          & 1.287     & 1.236   \\
A57    & Qantas Airways                & 4.342       & 4.949       & 4.696          & 4.561     & 4.527   \\
A58    & Qatar Airways                 & 4.899       & 4.612       & 4.561          & 4.460     & 4.831   \\
A59    & Rex Airlines                  & 1.523       & 1.473       & 1.574          & 1.101     & 1.253   \\
A60    & Royal Air Maroc               & 1.709       & 2.350       & 2.030          & 2.097     & 1.591   \\
A61    & Royal Brunei Airlines         & 2.620       & 2.857       & 2.350          & 2.418     & 2.688   \\
A62    & Ryanair                       & 3.397       & 3.262       & 3.734          & 3.329     & 3.329   \\
A63    & SAS Scandinavian              & 2.148       & 1.675       & 1.776          & 1.962     & 2.249   \\
A64    & Saudi Arabian Airlines        & 3.380       & 3.717       & 3.734          & 3.414     & 3.380   \\
A65    & Scoot                         & 3.397       & 4.038       & 3.802          & 3.582     & 3.295   \\
A66    & Singapore Airlines            & 4.848       & 4.831       & 4.511          & 4.477     & 4.629   \\
A67    & South African Airways         & 1.456       & 1.422       & 1.405          & 1.152     & 1.051   \\
A68    & Southwest Airlines            & 3.700       & 3.987       & 3.599          & 3.717     & 3.481   \\
A69    & SunExpress                    & 1.726       & 2.063       & 1.979          & 2.350     & 2.350   \\
A70    & Swiss International Air Lines & 4.646       & 4.612       & 4.544          & 4.781     & 4.578   \\
A71    & TAP Portugal                  & 2.097       & 1.861       & 1.793          & 1.928     & 2.013   \\
A72    & Thai Airways                  & 2.586       & 2.603       & 2.840          & 2.283     & 2.485   \\
A73    & Turkish Airlines              & 4.612       & 4.916       & 4.949          & 4.814     & 4.781   \\
A74    & United Airlines               & 1.776       & 1.692       & 1.759          & 2.215     & 1.895   \\
A75    & Vietnam Airlines              & 2.536       & 2.468       & 2.772          & 2.890     & 2.502   \\
A76    & Virgin Atlantic               & 4.257       & 4.190       & 4.713          & 3.684     & 3.869   \\
A77    & Virgin Australia              & 2.924       & 3.008       & 2.502          & 2.384     & 2.620   \\
A78    & Vistara                       & 3.987       & 4.359       & 3.954          & 3.869     & 4.173   \\
A79    & Vueling Airlines              & 3.008       & 2.570       & 2.468          & 3.278     & 3.177   \\
A80    & WestJet                       & 3.599       & 3.329       & 3.616          & 3.717     & 3.481   \\ \bottomrule
\end{tabular}}
\label{tab:table0101}
\end{table}

\begin{table}[pos=H]
\centering
\caption{Airline distances $D^{+}$, $D^{-}$ from positive and negative ideal solutions; positive and negative utility values $S^+$, $S^{-}$; positive and negative regret values $R^{+}$, $R^{-}$; harmonic mean $SDR^+$, $SDR^{-}$; airline ranking with corresponding $a_i$, $b_i$, $Q_i$.}
\resizebox{\linewidth}{!}{
{\begin{tabular}{p{2.5cm}p{6cm}lllllllllllp{2.5cm}}
\toprule
Number & Airline & $D^{+}$     & $D^{-}$     & $S^+$     & $S^{-}$     & $R^{+}$     & $R^{-}$  & $SDR^+$ & $SDR^-$   & $a$ & $b$ & $Q_i$ & Ranking \\ \midrule
A73                        & Turkish Airlines              & 1.80                 & 24.66 & 4.71                 & 95.29                & 0.97                 & 6.96                 & 2.49                 & 42.30                & 0.00                 & 0.00                 & 0.00                 & 1                    \\
A26                        & Emirates                      & 2.15                 & 24.31 & 6.17                 & 93.83                & 1.21                 & 7.17                 & 3.18                 & 41.77                & 0.02                 & 0.01                 & 0.02                 & 2                    \\
A48                        & Korean Air                    & 2.20                 & 24.21 & 6.44                 & 93.55                & 1.10                 & 6.91                 & 3.25                 & 41.56                & 0.02                 & 0.02                 & 0.02                 & 3                    \\
A42                        & Japan Airlines                & 2.66                 & 24.03 & 7.28                 & 92.72                & 1.59                 & 6.98                 & 3.84                 & 41.24                & 0.03                 & 0.03                 & 0.03                 & 4                    \\
A13                        & ANA All Nippon Airways        & 2.74                 & 23.87 & 7.92                 & 92.08                & 1.44                 & 6.98                 & 4.03                 & 40.98                & 0.04                 & 0.03                 & 0.04                 & 5                    \\
A58                        & Qatar Airways                 & 2.81                 & 23.80 & 8.22                 & 91.78                & 1.51                 & 7.05                 & 4.18                 & 40.88                & 0.04                 & 0.04                 & 0.04                 & 6                    \\
A6                         & Air France                    & 2.66                 & 23.71 & 8.40                 & 91.60                & 1.41                 & 7.07                 & 4.16                 & 40.80                & 0.04                 & 0.04                 & 0.04                 & 7                    \\
A66                        & Singapore Airlines            & 2.87                 & 23.62 & 8.74                 & 91.25                & 1.23                 & 6.91                 & 4.28                 & 40.59                & 0.05                 & 0.04                 & 0.04                 & 8                    \\
A70                        & Swiss International Air Lines & 3.15                 & 23.59 & 9.18                 & 90.82                & 1.85                 & 7.26                 & 4.73                 & 40.56                & 0.06                 & 0.04                 & 0.05                 & 9                    \\
A57                        & Qantas Airways                & 3.13                 & 23.50 & 9.54                 & 90.46                & 1.44                 & 6.98                 & 4.70                 & 40.31                & 0.06                 & 0.05                 & 0.05                 & 10                   \\
A47                        & KLM Royal Dutch Airlines      & 4.47                 & 21.81 & 15.99                & 84.01                & 2.08                 & 6.51                 & 7.51                 & 37.44                & 0.13                 & 0.12                 & 0.12                 & 11                   \\
A19                        & British Airways               & 4.92                 & 21.63 & 16.82                & 83.18                & 2.11                 & 6.50                 & 7.95                 & 37.10                & 0.14                 & 0.13                 & 0.13                 & 12                   \\
A23                        & China Southern Airlines       & 5.39                 & 21.52 & 17.77                & 82.23                & 2.59                 & 6.98                 & 8.58                 & 36.91                & 0.15                 & 0.14                 & 0.14                 & 13                   \\
A21                        & Cathay Pacific Airways        & 5.61                 & 21.34 & 18.29                & 81.71                & 2.57                 & 6.98                 & 8.82                 & 36.68                & 0.16                 & 0.14                 & 0.15                 & 14                   \\
A30                        & EVA Air                       & 5.69                 & 20.78 & 20.03                & 79.97                & 2.42                 & 6.51                 & 9.38                 & 35.75                & 0.17                 & 0.17                 & 0.17                 & 15                   \\
A40                        & Iberia                        & 5.97                 & 20.89 & 19.93                & 80.06                & 2.74                 & 6.32                 & 9.55                 & 35.76                & 0.18                 & 0.17                 & 0.17                 & 16                   \\
A24                        & Delta Air Lines               & 5.87                 & 20.77 & 20.21                & 79.79                & 2.66                 & 6.09                 & 9.58                 & 35.55                & 0.18                 & 0.17                 & 0.17                 & 17                   \\
A32                        & Finnair                       & 6.02                 & 20.64 & 20.76                & 79.24                & 2.73                 & 6.38                 & 9.84                 & 35.42                & 0.19                 & 0.17                 & 0.18                 & 18                   \\
A76                        & Virgin Atlantic               & 6.48                 & 20.32 & 21.95                & 78.04                & 2.83                 & 6.05                 & 10.42                & 34.81                & 0.20                 & 0.19                 & 0.19                 & 19                   \\
A28                        & Etihad Airways                & 6.34                 & 20.27 & 22.28                & 77.72                & 2.48                 & 6.23                 & 10.37                & 34.74                & 0.20                 & 0.19                 & 0.19                 & 20                   \\
A78                        & Vistara                       & 6.65                 & 19.86 & 23.67                & 76.33                & 2.75                 & 6.37                 & 11.02                & 34.19                & 0.22                 & 0.20                 & 0.21                 & 21                   \\
A18                        & Bangkok Airways               & 6.82                 & 19.88 & 23.98                & 76.02                & 2.59                 & 6.68                 & 11.13                & 34.20                & 0.22                 & 0.20                 & 0.21                 & 22                   \\
A36                        & Gulf Air                      & 6.87                 & 19.78 & 24.26                & 75.74                & 2.83                 & 6.68                 & 11.32                & 34.07                & 0.22                 & 0.21                 & 0.22                 & 23                   \\
A37                        & Hainan Airlines               & 7.01                 & 19.63 & 24.97                & 75.03                & 2.85                 & 6.15                 & 11.61                & 33.60                & 0.23                 & 0.22                 & 0.22                 & 24                   \\
A52                        & Lufthansa                     & 7.23                 & 19.48 & 25.62                & 74.38                & 2.93                 & 6.68                 & 11.93                & 33.51                & 0.24                 & 0.22                 & 0.23                 & 25                   \\
A68                        & Southwest Airlines            & 8.85                 & 17.44 & 32.99                & 67.01                & 3.19                 & 5.26                 & 15.01                & 29.90                & 0.32                 & 0.31                 & 0.31                 & 26                   \\
A9                         & AirAsia                       & 9.66                 & 17.16 & 34.84                & 65.15                & 3.99                 & 5.93                 & 16.16                & 29.42                & 0.34                 & 0.33                 & 0.34                 & 27                   \\
A65                        & Scoot                         & 9.83                 & 17.02 & 35.18                & 64.82                & 4.15                 & 5.40                 & 16.39                & 29.08                & 0.35                 & 0.33                 & 0.34                 & 28                   \\
A14                        & Asiana Airlines               & 9.88                 & 16.66 & 36.57                & 63.42                & 3.41                 & 5.37                 & 16.62                & 28.49                & 0.36                 & 0.35                 & 0.35                 & 29                   \\
A80                        & WestJet                       & 10.09                & 16.75 & 36.62                & 63.38                & 4.16                 & 5.40                 & 16.96                & 28.51                & 0.36                 & 0.35                 & 0.36                 & 30                   \\
A64                        & Saudi Arabian Airlines        & 10.23                & 16.64 & 37.17                & 62.83                & 4.21                 & 5.93                 & 17.20                & 28.47                & 0.37                 & 0.35                 & 0.36                 & 31                   \\
A31                        & Fiji Airways                  & 10.09                & 16.41 & 37.63                & 62.37                & 3.45                 & 5.40                 & 17.06                & 28.06                & 0.37                 & 0.36                 & 0.36                 & 32                   \\
A15                        & Austrian Airlines             & 10.31                & 16.34 & 37.67                & 62.33                & 4.43                 & 5.40                 & 17.47                & 28.02                & 0.38                 & 0.36                 & 0.37                 & 33                   \\
A27                        & Ethiopian Airlines            & 10.52                & 15.89 & 39.17                & 60.83                & 3.90                 & 5.20                 & 17.86                & 27.31                & 0.39                 & 0.38                 & 0.38                 & 34                   \\
A7                         & Air New Zealand               & 10.67                & 15.69 & 40.17                & 59.83                & 3.87                 & 5.40                 & 18.24                & 26.97                & 0.40                 & 0.39                 & 0.39                 & 35                   \\
A62                        & Ryanair                       & 10.68                & 15.71 & 40.10                & 59.90                & 4.08                 & 5.17                 & 18.29                & 26.93                & 0.40                 & 0.39                 & 0.39                 & 36                   \\
A1                         & Aegean Airlines               & 11.07                & 15.55 & 41.00                & 59.00                & 4.12                 & 5.31                 & 18.73                & 26.62                & 0.41                 & 0.40                 & 0.40                 & 37                   \\
A4                         & Air Astana                    & 11.19                & 15.34 & 41.82                & 58.17                & 4.16                 & 5.40                 & 19.06                & 26.30                & 0.42                 & 0.40                 & 0.41                 & 38                   \\
A54                        & Oman Air                      & 11.67                & 14.67 & 43.94                & 56.06                & 3.90                 & 4.92                 & 19.84                & 25.22                & 0.44                 & 0.43                 & 0.43                 & 39                   \\
A35                        & Garuda Indonesia              & 11.64                & 14.75 & 43.95                & 56.05                & 4.06                 & 4.96                 & 19.88                & 25.25                & 0.44                 & 0.43                 & 0.43                 & 40                   \\
A79                        & Vueling Airlines              & 13.85                & 12.53 & 52.78                & 47.22                & 4.57                 & 4.08                 & 23.74                & 21.28                & 0.54                 & 0.53                 & 0.53                 & 41                   \\
A44                        & JetBlue Airways               & 14.37                & 12.38 & 53.95                & 46.05                & 5.22                 & 4.13                 & 24.52                & 20.85                & 0.56                 & 0.54                 & 0.55                 & 42                   \\
A5                         & Air Canada                    & 14.77                & 11.70 & 56.35                & 43.65                & 4.91                 & 4.25                 & 25.34                & 19.87                & 0.58                 & 0.57                 & 0.57                 & 43                   \\
A17                        & Azul Airlines                 & 14.76                & 11.59 & 56.40                & 43.60                & 4.78                 & 4.25                 & 25.31                & 19.81                & 0.58                 & 0.57                 & 0.57                 & 44                   \\
A25                        & EasyJet                       & 14.76                & 11.58 & 56.40                & 43.59                & 5.01                 & 4.13                 & 25.39                & 19.77                & 0.58                 & 0.57                 & 0.57                 & 45                   \\
A53                        & Malaysia Airlines             & 15.09                & 11.53 & 57.08                & 42.92                & 5.40                 & 4.25                 & 25.85                & 19.57                & 0.59                 & 0.57                 & 0.58                 & 46                   \\
A22                        & China Airlines                & 15.31                & 11.32 & 57.71                & 42.29                & 5.85                 & 4.08                 & 26.29                & 19.23                & 0.60                 & 0.58                 & 0.59                 & 47                   \\
A77                        & Virgin Australia              & 15.32                & 11.25 & 58.22                & 41.78                & 5.17                 & 4.25                 & 26.24                & 19.09                & 0.60                 & 0.59                 & 0.59                 & 48                   \\
A75                        & Vietnam Airlines              & 15.80                & 10.91 & 59.75                & 40.25                & 5.50                 & 3.98                 & 27.01                & 18.38                & 0.62                 & 0.60                 & 0.61                 & 49                   \\
A50                        & LATAM                         & 15.85                & 10.74 & 60.27                & 39.73                & 5.28                 & 3.89                 & 27.13                & 18.12                & 0.62                 & 0.61                 & 0.62                 & 50                   \\
A61                        & Royal Brunei Airlines         & 16.07                & 10.48 & 61.02                & 38.98                & 5.28                 & 3.98                 & 27.46                & 17.81                & 0.63                 & 0.62                 & 0.62                 & 51                   \\
A2                         & Aer Lingus                    & 16.35                & 10.31 & 61.84                & 38.16                & 5.40                 & 4.08                 & 27.86                & 17.51                & 0.64                 & 0.63                 & 0.63                 & 52                   \\
A72                        & Thai Airways                  & 16.39                & 10.19 & 61.99                & 38.01                & 5.85                 & 3.55                 & 28.08                & 17.25                & 0.65                 & 0.63                 & 0.64                 & 53                   \\
A16                        & Azerbaijan Airlines           & 16.61                & 9.75  & 63.66                & 36.34                & 5.32                 & 3.88                 & 28.53                & 16.66                & 0.66                 & 0.65                 & 0.65                 & 54                   \\
A41                        & IndiGo                        & 16.67                & 9.68  & 63.51                & 36.49                & 5.66                 & 3.61                 & 28.61                & 16.59                & 0.66                 & 0.65                 & 0.65                 & 55                   \\
A34                        & Flynas                        & 18.63                & 7.58  & 71.66                & 28.34                & 5.68                 & 2.57                 & 31.99                & 12.83                & 0.74                 & 0.74                 & 0.74                 & 56                   \\
A39                        & Hong Kong Airlines            & 18.85                & 7.48  & 72.40                & 27.60                & 5.69                 & 3.17                 & 32.31                & 12.75                & 0.75                 & 0.75                 & 0.75                 & 57                   \\
A45                        & Jetstar Airways               & 18.98                & 7.51  & 72.81                & 27.19                & 5.81                 & 2.89                 & 32.53                & 12.53                & 0.76                 & 0.75                 & 0.75                 & 58                   \\
A69                        & SunExpress                    & 19.16                & 7.43  & 73.34                & 26.65                & 6.78                 & 2.83                 & 33.09                & 12.30                & 0.77                 & 0.76                 & 0.76                 & 59                   \\
A43                        & Jet2.com                      & 19.56                & 7.43  & 74.27                & 25.73                & 6.98                 & 2.89                 & 33.60                & 12.01                & 0.78                 & 0.76                 & 0.77                 & 60                   \\
A38                        & Hawaiian Airlines             & 19.51                & 7.15  & 74.68                & 25.32                & 6.50                 & 2.83                 & 33.56                & 11.77                & 0.78                 & 0.77                 & 0.78                 & 61                   \\
A29                        & Eurowings                     & 19.87                & 6.58  & 76.26                & 23.74                & 6.33                 & 2.84                 & 34.15                & 11.05                & 0.80                 & 0.79                 & 0.79                 & 62                   \\
A63                        & SAS Scandinavian              & 19.94                & 6.47  & 76.64                & 23.36                & 6.03                 & 2.71                 & 34.20                & 10.85                & 0.80                 & 0.79                 & 0.80                 & 63                   \\
A71                        & TAP Portugal                  & 20.09                & 6.64  & 77.07                & 22.93                & 6.14                 & 2.84                 & 34.43                & 10.80                & 0.81                 & 0.80                 & 0.80                 & 64                   \\
A60                        & Royal Air Maroc               & 20.15                & 6.48  & 77.06                & 22.94                & 6.51                 & 2.76                 & 34.57                & 10.73                & 0.81                 & 0.80                 & 0.80                 & 65                   \\
A11                        & Alaska Airlines               & 20.42                & 5.99  & 78.61                & 21.39                & 6.51                 & 2.46                 & 35.18                & 9.95                 & 0.82                 & 0.82                 & 0.82                 & 66                   \\
A74                        & United Airlines               & 20.66                & 6.12  & 79.10                & 20.90                & 6.31                 & 2.83                 & 35.36                & 9.95                 & 0.83                 & 0.82                 & 0.82                 & 67                   \\
A12                        & American Airlines             & 20.68                & 6.12  & 79.37                & 20.63                & 6.29                 & 2.83                 & 35.44                & 9.86                 & 0.83                 & 0.82                 & 0.83                 & 68                   \\
A10                        & airBaltic                     & 20.95                & 6.03  & 79.95                & 20.05                & 6.50                 & 2.60                 & 35.80                & 9.56                 & 0.84                 & 0.83                 & 0.83                 & 69                   \\
A55                        & Peach                         & 20.92                & 5.83  & 80.04                & 19.96                & 6.78                 & 2.51                 & 35.91                & 9.43                 & 0.84                 & 0.83                 & 0.84                 & 70                   \\
A8                         & Air Transat                   & 23.17                & 3.41  & 89.38                & 10.62                & 6.74                 & 1.58                 & 39.76                & 5.21                 & 0.94                 & 0.94                 & 0.94                 & 71                   \\
A51                        & LOT Polish                    & 23.16                & 3.20  & 89.36                & 10.63                & 7.07                 & 1.38                 & 39.86                & 5.07                 & 0.94                 & 0.94                 & 0.94                 & 72                   \\
A46                        & Jetstar Asia                  & 23.34                & 3.11  & 90.05                & 9.95                 & 6.88                 & 1.56                 & 40.09                & 4.87                 & 0.95                 & 0.94                 & 0.95                 & 73                   \\
A56                        & Philippine Airlines           & 23.74                & 2.89  & 91.48                & 8.52                 & 7.26                 & 1.56                 & 40.83                & 4.32                 & 0.97                 & 0.96                 & 0.96                 & 74                   \\
A59                        & Rex Airlines                  & 23.82                & 2.95  & 91.55                & 8.45                 & 7.26                 & 1.42                 & 40.88                & 4.27                 & 0.97                 & 0.96                 & 0.96                 & 75                   \\
A33                        & flyDubai                      & 23.82                & 2.86  & 91.75                & 8.25                 & 7.26                 & 1.58                 & 40.94                & 4.23                 & 0.97                 & 0.96                 & 0.97                 & 76                   \\
A49                        & Kuwait Airways                & 23.93                & 2.74  & 92.39                & 7.61                 & 6.91                 & 1.43                 & 41.07                & 3.93                 & 0.97                 & 0.97                 & 0.97                 & 77                   \\
A20                        & Brussels Airlines             & 23.95                & 2.62  & 92.45                & 7.54                 & 7.05                 & 1.42                 & 41.15                & 3.86                 & 0.98                 & 0.97                 & 0.97                 & 78                   \\
A67                        & South African Airways         & 24.31                & 2.43  & 93.63                & 6.36                 & 7.07                 & 1.46                 & 41.67                & 3.42                 & 0.99                 & 0.98                 & 0.99                 & 79                   \\
A3                         & Air Arabia                    & 24.56                & 1.87  & 94.92                & 5.08                 & 6.90                 & 1.13                 & 42.13                & 2.69                 & 1.00                 & 1.00                 & 1.00                 & 80                   \\ \bottomrule
\end{tabular}}}
\label{tab:table1}
\end{table}

\begin{table}[pos=H]
\centering
\caption{Ranking of airline service quality.}
\begin{tabular}{lllllll}
\toprule
\multicolumn{1}{c}{Ranking} & \multicolumn{1}{c}{Number} & \multicolumn{1}{c}{Airline}   & \multicolumn{1}{c}{} & \multicolumn{1}{c}{Ranking} & \multicolumn{1}{c}{Number} & \multicolumn{1}{c}{Airline} \\ \midrule
=1                          & A26                        & Emirates                      &                      & =15                         & A5                         & Air Canada                  \\
=1                         & A73                        & Turkish Airlines              &
& =15                         & A17                        & Azul Airlines               \\
1-2                          & A70                        & Swiss International Air Lines &
& 15-16                       & A44                        & JetBlue Airways             \\
=2                          & A6                         & Air France                    &                      & 15-16                       & A53                        & Malaysia Airlines           \\
=2                          & A13                        & ANA All Nippon Airways        &                      & 15-17                       & A77                        & Virgin Australia            \\
=2                          & A42                        & Japan Airlines                &                      & 15-17                       & A79                        & Vueling Airlines            \\
=2                          & A48                        & Korean Air                    &                      & 16                          & A25                        & EasyJet                     \\
=2                          & A58                        & Qatar Airways                 &                      & 17                          & A2                         & Aer Lingus                  \\
=3                          & A21                        & Cathay Pacific Airways        &                      & 17-19                       & A22                        & China Airlines              \\
=3                          & A23                        & China Southern Airlines       &                      & 18                          & A61                        & Royal Brunei Airlines       \\
3-4                         & A57                        & Qantas Airways                &                      & 18-19                       & A75                        & Vietnam Airlines            \\
3-4                         & A66                        & Singapore Airlines            &                      & 19                          & A50                        & LATAM                       \\
4                           & A18                        & Bangkok Airways               &                      & =20                         & A16                        & Azerbaijan Airlines         \\
4-5                         & A30                        & EVA Air                       &                      & =20                         & A41                        & IndiGo                      \\
4-7                         & A36                        & Gulf Air                      &                      & =20                         & A72                        & Thai Airways                \\
4-7                         & A47                        & KLM Royal Dutch Airlines      &                      & 21-25                       & A34                        & Flynas                      \\
4-8                         & A52                        & Lufthansa                     &                      & 21                          & A39                        & Hong Kong Airlines          \\
5                           & A19                        & British Airways               &                      & 21-22                       & A45                        & Jetstar Airways             \\
=6                          & A32                        & Finnair                       &                      & 22-22                       & A43                        & Jet2.com                    \\
=6                          & A40                        & Iberia                        &                      & 22-23                       & A69                        & SunExpress                  \\
=7                          & A24                        & Delta Air Lines               &                      & =23                         & A29                        & Eurowings                   \\
=7                          & A28                        & Etihad Airways                &                      & =23                         & A38                        & Hawaiian Airlines           \\
=7                          & A78                        & Vistara                       &                      & =23                         & A71                        & TAP Portugal                \\
=8                          & A37                        & Hainan Airlines               &                      & =24                         & A12                        & American Airlines           \\
=8                          & A76                        & Virgin Atlantic               &                      & =24                         & A60                        & Royal Air Maroc             \\
=9                          & A9                         & AirAsia                       &                      & =24                         & A63                        & SAS Scandinavian            \\
=9                          & A64                        & Saudi Arabian Airlines        &                      & =24                         & A74                        & United Airlines             \\
9-11                        & A68                        & Southwest Airlines            &                      & =25                         & A10                        & airBaltic                   \\
=10                         & A31                        & Fiji Airways                  &                      & =25                         & A11                        & Alaska Airlines             \\
=10                         & A65                        & Scoot                         &                      & 26                          & A55                        & Peach                       \\
=10                         & A80                        & WestJet                       &                      & 27                          & A8                         & Air Transat                 \\
=11                         & A4                         & Air Astana                    &                      & 27-28                       & A33                        & flyDubai                    \\
=11                         & A7                         & Air New Zealand               &                      & 27-30                       & A51                        & LOT Polish                  \\
11-12                       & A14                        & Asiana Airlines               &                      & =28                         & A46                        & Jetstar Asia                \\
11-13                       & A15                        & Austrian Airlines             &                      & =28                         & A56                        & Philippine Airlines         \\
12-13                       & A1                         & Aegean Airlines               &                      & 29                          & A49                        & Kuwait Airways              \\
12                          & A27                        & Ethiopian Airlines            &                      & 29-30                       & A59                        & Rex Airlines                \\
13                          & A62                        & Ryanair                       &                      & 29-30                       & A67                        & South African Airways       \\
=14                         & A35                        & Garuda Indonesia              &                      & =30                         & A3                         & Air Arabia                  \\
=14                         & A54                        & Oman Air                      &                      & =30                         & A20                        & Brussels Airlines            \\\bottomrule
\end{tabular}
\label{tab:table1010}
\end{table}

\subsection{Sensitivity analysis}

\begin{figure}[pos=H]
\centering
\includegraphics[width=0.7\linewidth]{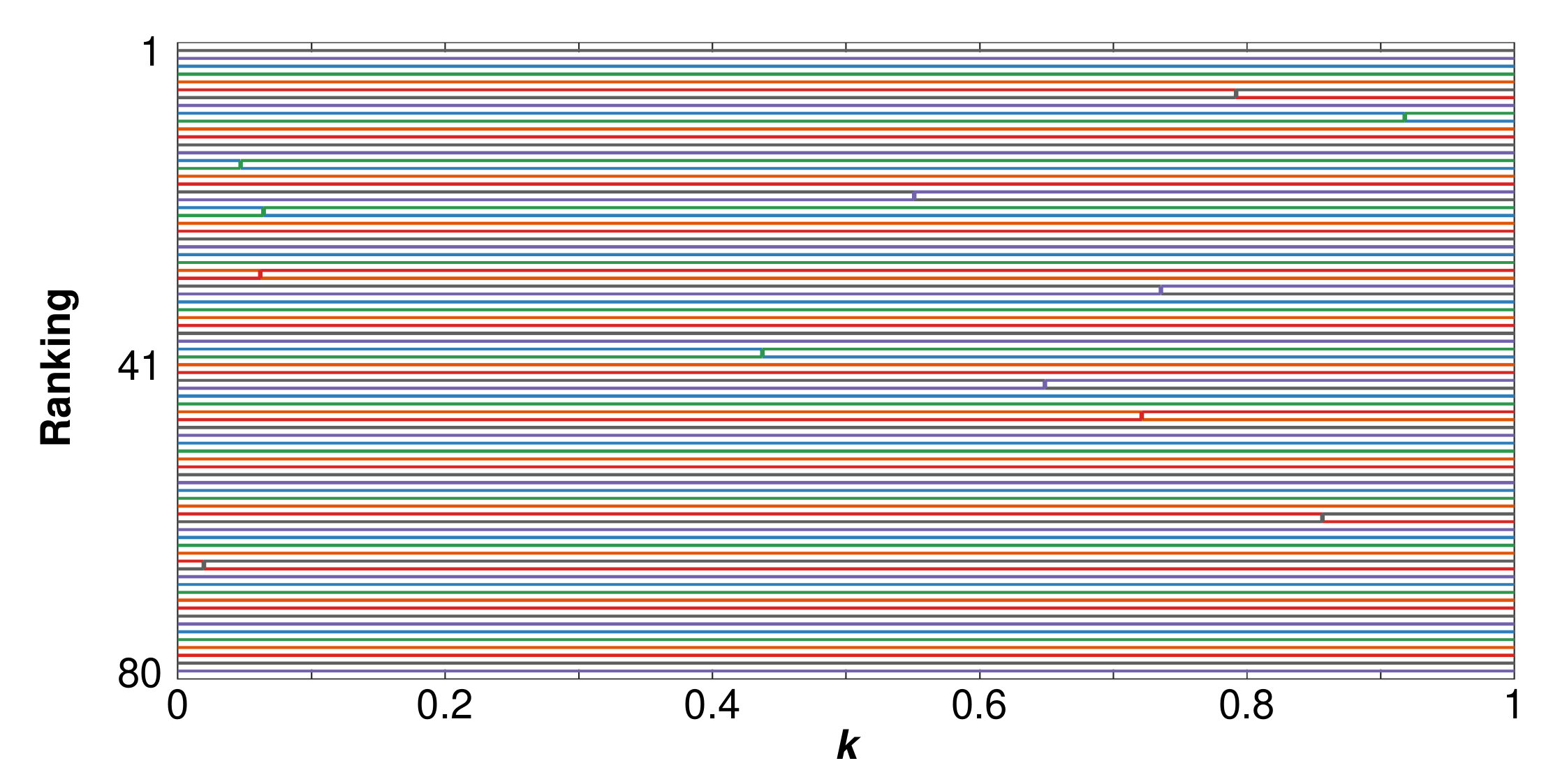}
\caption {Results of sensitivity analysis. The fluctuation of the curve shows how the change in $k$ affects the ranking of airline service quality.}
\label{fig:figure4}
\end{figure}

To evaluate the robustness of the proposed integrated methodology, we conducted a sensitivity analysis. The value of the allocation coefficient $k$ may have an impact on how $a$ and $b$ are allocated, as shown in equation (\ref{eq:equation16}). Adjustments to $k$ result in varying service quality scores for each airline, thus affecting their ranking. We methodically altered the value of $k$ across a continuum from 0 to 1, in increments of 0.01. The outcomes of this sensitivity analysis are depicted in Figure \ref{fig:figure4}, with the horizontal axis illustrating the alterations in $k$ and the vertical axis representing the 80 airlines.

The analysis reveals that the trajectories representing the variation in the 80 $k$ values predominantly exhibit stability, with minor fluctuations. Notably, the few exceptions deviate by a marginal range, not exceeding one unit. These observations decisively indicate that the results derived from the proposed methodology demonstrate a high degree of reliability and are conducive for decision-making in this context.

\section{Discussion}
\label{Discussion}

In our research, we applied text mining and Multi-Criteria Decision Making (MCDM) techniques to online reviews of 80 airlines, uncovering intriguing patterns and insights into customer perceptions and preferences regarding airline service quality. Our proposed methodology facilitates a multi-dimensional satisfaction assessment, demonstrating that more effective and realistic evaluations and analyses are achievable. The case study findings highlight that passengers place significant emphasis on aspects such as aircraft safety, service provision, and staff conduct, expressing satisfaction or dissatisfaction with these elements of service quality. Commonly occurring terms in the reviewed online feedback, such as "flight", "service", "staff", "seat", "food", and "time", suggest that airlines and stakeholders should intensify their focus on these service quality facets to enhance passenger satisfaction \citep{park2020understanding}.

Utilizing our proposed method, the comparative ranking of service quality between any two airlines can be determined through a series of directed topological hierarchical graphs, as depicted in Figures\ref{fig:figure2} and \ref{fig:figure3}. This framework constitutes a rigid structure. Within these graphs, an element is designated as active if it occupies a distinct topological level. Systems comprising active elements are termed extensionally variable systems, in contrast to topologically rigid systems that lack such elements. The inherent stability and robustness of relationships and order within a topologically rigid system are notable. Based on our analysis, the rankings generated by this method can be deemed stable, accurately reflecting the comparative service quality across various airlines.

Compared to other traditional MCDM models, our comprehensive model offers distinct advantages. It not only incorporates text mining as a data source but also delineates hierarchies among the evaluated alternatives. This hierarchical structure facilitates the evaluation of entities whose relative rankings may not be initially apparent. By employing this method, stable rankings can be derived in diverse fields, such as the hotel industry \citep{berezina2016understanding} and the supply chain sector \citep{lim2021exploring}.

Nonetheless, it is vital to acknowledge the limitations and challenges of these methodologies. A notable constraint of text mining is its potential inability to fully grasp the subtleties and context of natural language, including sarcasm and humor \citep{gupta2020comprehensive,kumar2021applications,ashtiani2023news}, which could impact the precision and efficacy of topic and sentiment extraction. Additionally, customer subjectivity and biases, influenced by personal expectations, emotions, or factors \citep{hassani2020text,kushwaha2021applications}, might skew the objective representation of service quality. Therefore, a cautious and critical approach is essential in interpreting and generalizing these findings to attain a more comprehensive and reliable understanding of airline service quality.

\section{Conclusion}
\label{Conclusion}

In this study, we have devised a comprehensive model for assessing airline service quality, utilizing the LSA-TOPSIS-VIKOR-AISM methodology. This comprehensive framework commences with the extraction of online review data from various platforms, subsequently assigning sentiment scores to each review. Our analysis delves into the utility, Euclidean distance, and individual regret metrics for each airline, culminating in the derivation of ranked trade-off solutions. Furthermore, we methodically scrutinize various indicators using the AISM approach, facilitating the creation of directed topological hierarchical graphs. The empirical evidence suggests that our proposed method boasts of robust data integrity, streamlined computations, and cogent conclusions, making it an exemplary tool for a holistic evaluation of service quality or performance.

Our research illuminates the fusion of burgeoning technologies—text mining and sentiment analysis—with the MCDM model, harnessing the strengths of both to enhance airline service quality evaluation, and effectively captures consumer intentions in a comprehensive manner. Future inquiries might expand to encompass a broader array of airlines, a wider range of service indicators, more extensive data sets, or the amalgamation with other methodologies, such as machine learning \citep{liao2023reimagining} and complex networks \citep{karcz2021multi}. While the distribution coefficient $k$ used in our trade-off solution calculations is reflective of specific airline professionals' opinions, it may not entirely align with the general public's preferences in airline decision-making. The incorporation of prospect theory \citep{zhao2022selecting} and the construction of a value function based on multiple mental accounts opens avenues for future research into the variability of $k$ under diverse scenarios.

\section*{Data collection protocol}
The data supporting the findings of this study are available from the corresponding author upon request.

\section*{Acknowledgements}
The authors extend their sincere gratitude to Mr. Zhigang Wang for providing invaluable suggestions.

\printcredits

\bibliographystyle{cas-model2-names}

\bibliography{cas-refs}


\end{document}